\documentclass{llncs}
\usepackage{llncsdoc}
\usepackage[english]{babel}
\usepackage[latin1]{inputenc}
\usepackage{amssymb}
\usepackage{amsfonts}
\usepackage{mathrsfs}
\usepackage{amsmath}
\usepackage[dvips]{graphicx}
\usepackage{graphicx,graphics}
\usepackage{rotate}
\usepackage{epsfig,bm}

\begin{document}

\title{Dynamical localization in kicked rotator as a paradigm of other systems: spectral statistics and the localization measure}

\author{Thanos Manos\inst{1,2} \and Marko Robnik\inst{1}}

\institute{University of Maribor,
Center for Applied Mathematics and Theoretical Physics, \\Krekova 2,
SI-2000 Maribor, SLOVENIA
\and
University of Nova Gorica, School of applied sciences, \\Vipavska 11c,
SI-5270, Ajdov\v{s}\v{c}ina, SLOVENIA}

\maketitle

\begin{abstract}
We study the intermediate statistics of the spectrum of quasi-energies and of the eigenfunctions in the kicked rotator, in the case when the corresponding system is fully chaotic while quantally localized. As for the eigenphases, we find clear evidence that the spectral statistics is well described by the Brody distribution, notably better than by the Izrailev's one, which has been proposed and used broadly to describe such cases. We also studied the eigenfunctions of the Floquet operator and their localization. We show the existence of a scaling law between the repulsion parameter with relative localization length, but only as a first order approximation, since another parameter plays a role. We believe and have evidence that a similar analysis applies in time-independent Hamilton systems.
\end{abstract}

\section{Introduction}\label{sec:intro}

One of the most important manifestations of quantum chaos  of low-dimensional classically fully chaotic (ergodic) Hamiltonian systems is the fact that in the (sufficiently deep) semiclassical limit the statistical properties of the discrete energy spectra obey the statistics of Gaussian Random Matrix Theory (RMT). The opposite extreme are classically integrable systems, which quantally exhibit Poissonian spectral statistics (see \cite{HaakeBook2001}).

Quantum kicked rotator (QKR) is a typical example in the field of quantum chaos \cite{StockBook:1999}. A typical property of the QKR is the chaos suppression for sufficiently large time scales. The study of the statistical properties of the classical and quantum (semiclassical) parameters in such systems is of great importance. Here we study in detail the semiclassical region where $k>K>1$, i.e. the regime of full correspondence between quantum and classical diffusion (on the finite time scale $t \leq t_D$) and the manifested quantum dynamical localization for $t > t_D$. Furthermore, we are focused in the probability level spacing distributions in the regime where the system is classically strongly chaotic ($K \geq 7$) but quantally localized, i.e in the \textit{intermediate} or \textit{soft} quantum chaos, as it is described in the literature \cite{IZ_PR:1990}.

\section{The quantum kicked rotator model}\label{sec:QKR}
The QKR model \cite{CCIF1979} is described by the following function
\begin{equation}\label{QSM}
\hat{H}=-\frac{\hbar^2}{2I}\frac{\partial^2}{\partial \theta^2}+\varepsilon_0 \cos \theta \sum_{m=-\infty}^{\infty} \delta(t-mT),
\end{equation}
where $\hbar$ is Planck's constant, $I$ is the moment of inertia of the pendulum and $\varepsilon_0$ is the perturbation strength. The motion after one period $T$ of the $\psi$ wave function then can be described by the following mapping
\begin{flalign}
  &\psi(\theta,t+T) = \hat{U}\psi(\theta,t), \\
  &\hat{U} = \exp \left ( i \frac{T\hbar}{4I}\frac{\partial^2}{\partial \theta^2} \right )\exp \left (-i\frac{\varepsilon_0}{\hbar} \cos \theta \right)\exp \left (i \frac{T\hbar}{4I}\frac{\partial^2}{\partial \theta^2} \right),
\end{flalign}
where the $\psi$ function is determined in the middle of the rotation, between two successive kicks. The evolution operator $\hat{U}$ of the system corresponds to one period. Due to the instant action of the perturbation, this evolution can be written as the product of three non-commuting unitary operators, the first and third of which corresponds to the free rotation during half a period $ \hat{G}(\tau/2)=\exp \left (i\frac{T\hbar}{4I}\frac{\partial^2}{\partial \theta^2} \right)$, $ \tau \equiv \hbar T/I$, while the second $\hat{B}(k)=\exp(-ik\cos \theta)$, $ k \equiv \varepsilon_0/\hbar$ describes the kick. The system's behavior depends only on two parameters, i.e. $\tau$ and $k$ and its correspondence with the classical systems is described by the relation $K=k \tau=\varepsilon_0 T/I$. In the case $K\equiv k \tau >>1$ the motion is well-known to be strongly chaotic. The transition to classical mechanics is described by the limit $k \rightarrow \infty$, $\tau \rightarrow 0$ while $K=\text{const}$. In what follows $\hbar=\tau$ and $T=I=1$. We shall consider mostly the semiclassical regime $k>K$, where $\tau <1$.

In order to study how the localization affects the statistical properties of the quasienergy spectra we use the model's representation with a finite number $N$ of levels \cite{IZ_PLA:1988,IZ_PR:1990}
\begin{equation}\label{psi_repres}
\psi_n(t+T) = \sum_{m=1}^{N} U_{nm}\psi_m(t), \qquad n,m=1,2,...,N \enspace .
\end{equation}
The finite unitary matrix $\hat{U}_{nm}$ determines the evolution of a $N$-dimensional vector (Fourier transform of $\psi$) of the model
\begin{equation}\label{Unm}
    U_{nm}=\sum_{n'm'}G_{nm'}B_{n'm'}G_{n'm},
\end{equation}
where $G_{ll'}=\exp \left (i\tau l^2/4 \right )\delta_{ll'}$ is a diagonal matrix corresponding to free rotation during a half period $T/2$ and  the matrix $B_{n'm'}$ describing the one kick has the following form
\begin{flalign}\label{Bnmoper}
  B_{n'm'}&= \frac{1}{2N+1}\sum_{l=1}^{2N+1} \left \{ \cos \left [ \left (n'-m' \right ) \frac{2 \pi l}{2N+1}\right ] - \cos \left [(n'+m')\frac{2 \pi l}{2N+1} \right ] \right \} \notag\\ \nonumber
  & \times \exp \left [-ik\cos \left (\frac{2 \pi l}{2N+1}\right ) \right ]. \nonumber
\end{flalign}
The model (\ref{psi_repres}) with a finite number of states is considered as the quantum analogue of the classical standard mapping on the torus with closed momentum $p$ and phase $\theta$ where $U_{mn}$ describes only the odd states of the systems, i.e. $\psi(\theta)=-\psi(-\theta)$.

\section{Intermediate statistics and comparison of probability distributions}\label{sec:interstat}

Let us first compare the Brody and Izrailev probability distribution functions (PDFs) for the study of the intermediate level statistics. The Brody distribution is defined by the relation
\begin{equation}\label{BRpdf}
    P_{\textrm{BR}}(s)=C_1 s^{\beta_{\textrm{BR}}} \exp \left (-C_2 s^{\beta_\textrm{BR}+1} \right ),
\end{equation}
where the two parameters $C_1$ and $C_2$ are determined by the normalization conditions $\int_0^\infty P_{\textrm{BR}}(s) ds=1$ and $\int_0^\infty s P_B(s) ds=1$. Izrailev suggested the following distribution (see \cite{IZ_PR:1990} and references there for the details and the argumentation)
\begin{equation}\label{IZpdf}
P_{\textrm{IZ}}(s) = A\left (\frac{1}{2}\pi s \right )^{\beta_\textrm{IZ}} \exp \left [-\frac{1}{16}\beta_\textrm{IZ} \pi^2 s^2 -\left (B-\frac{1}{4}\pi \beta_\textrm{IZ} \right )s \right],
\end{equation} in order to describe the intermediate statistics, where the  parameters $A$ and $B$ are determined again by the two above normalization conditions. Both $\beta$ parameters, in the strongly localized regime tend asymptotically to 0 with Poissonian statistics while in the chaotic one tend to 1, which excellently approximates the Gaussian Orthogonal Ensemble (GEO) of the RMT. On the other hand, the non-integer $\beta$ in the PDFs could be associated with the statistics of the quasienergy states with chaotic localized eigenfunctions.
\begin{figure}[h]
\centering
\includegraphics[width=6.0cm]{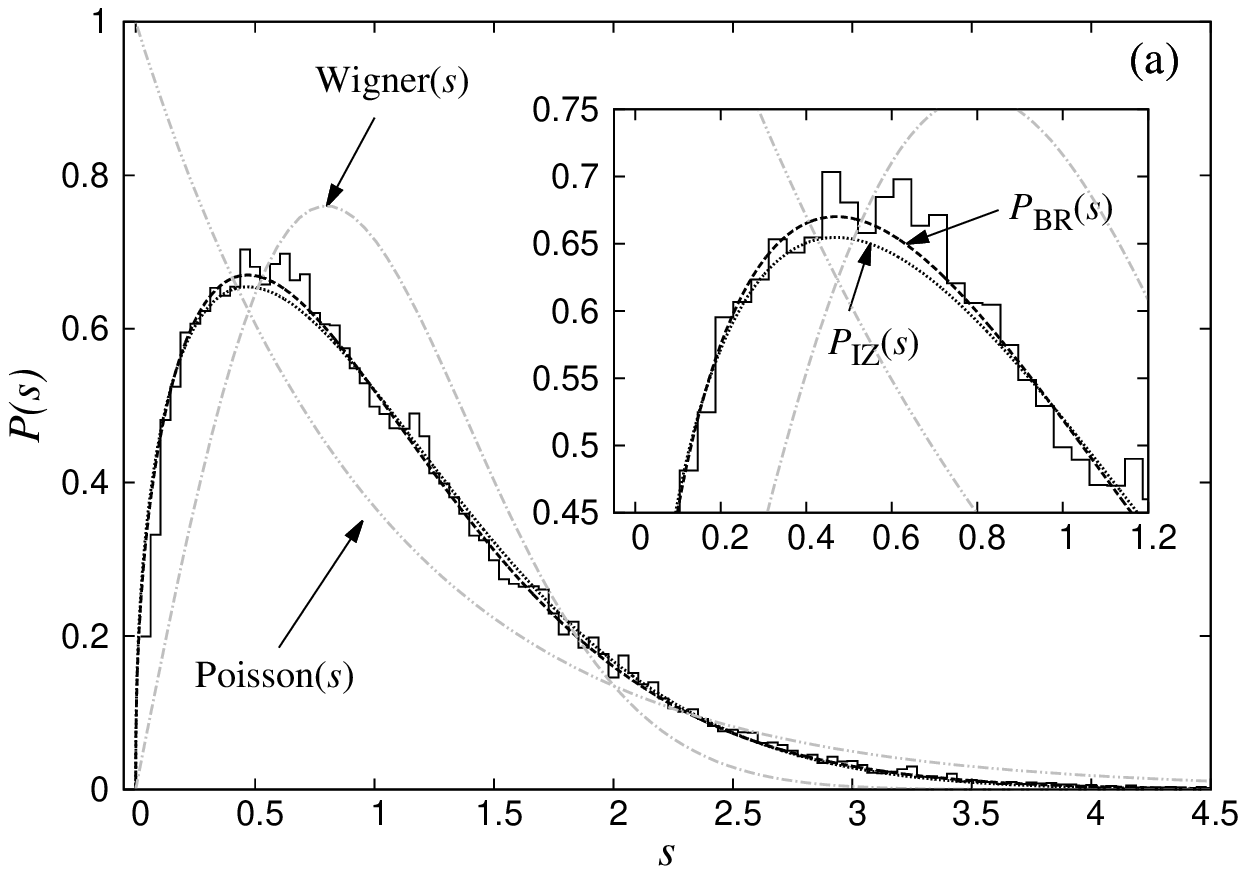}
\includegraphics[width=6.0cm]{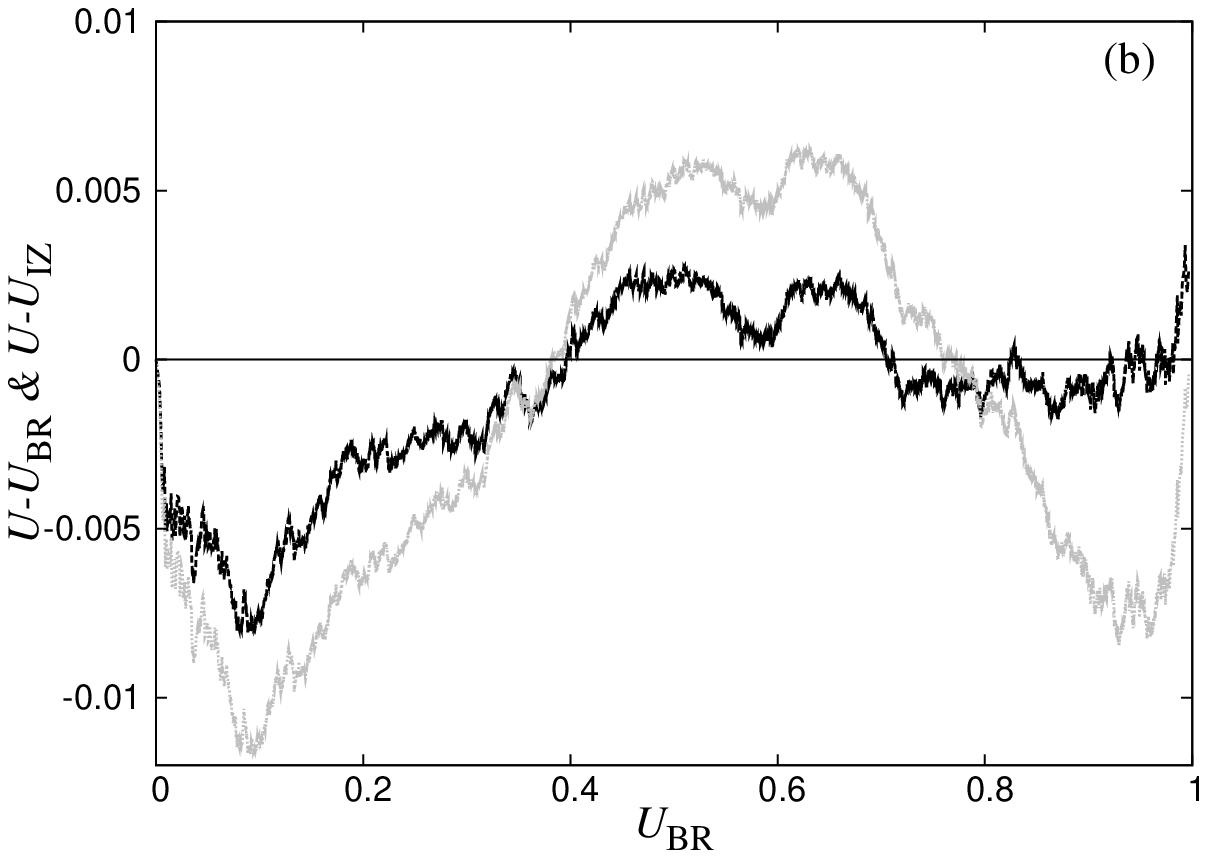}\\
\includegraphics[width=6.0cm]{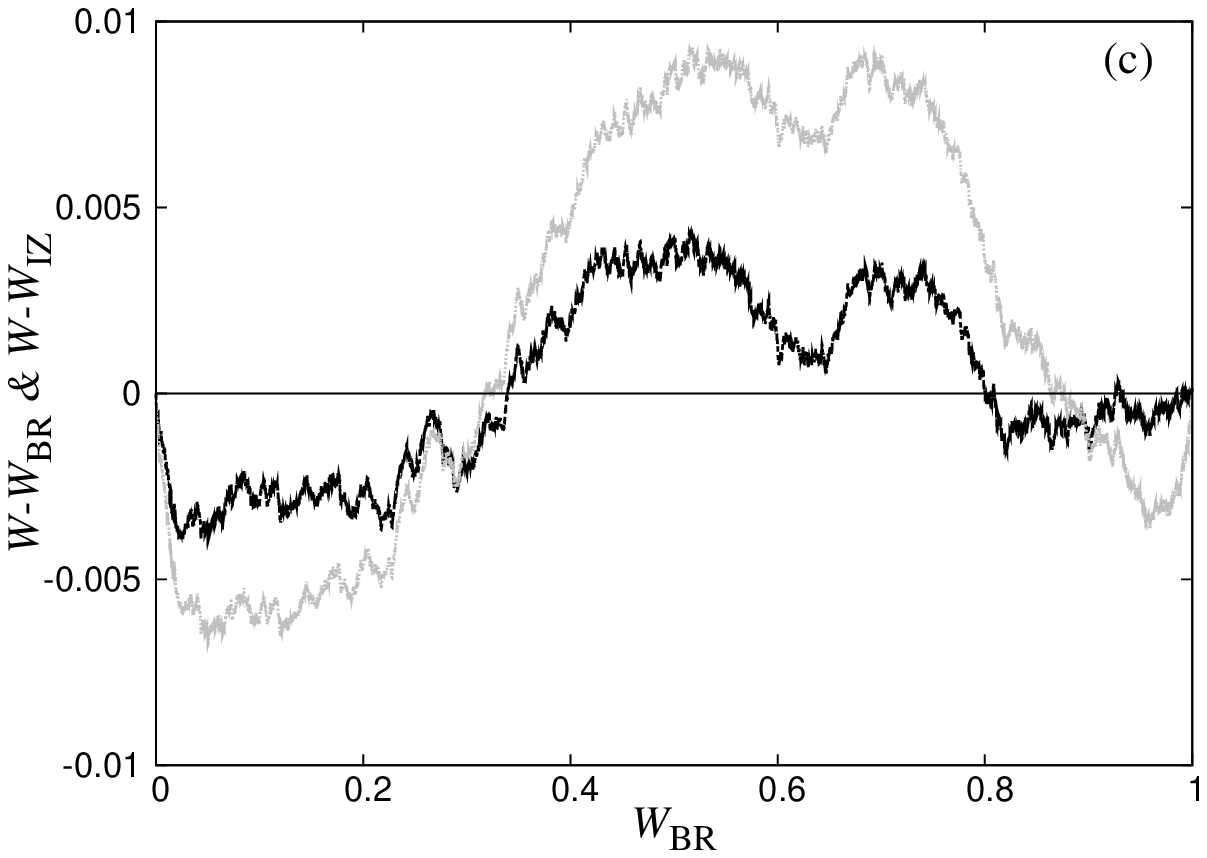}
\includegraphics[width=6.0cm]{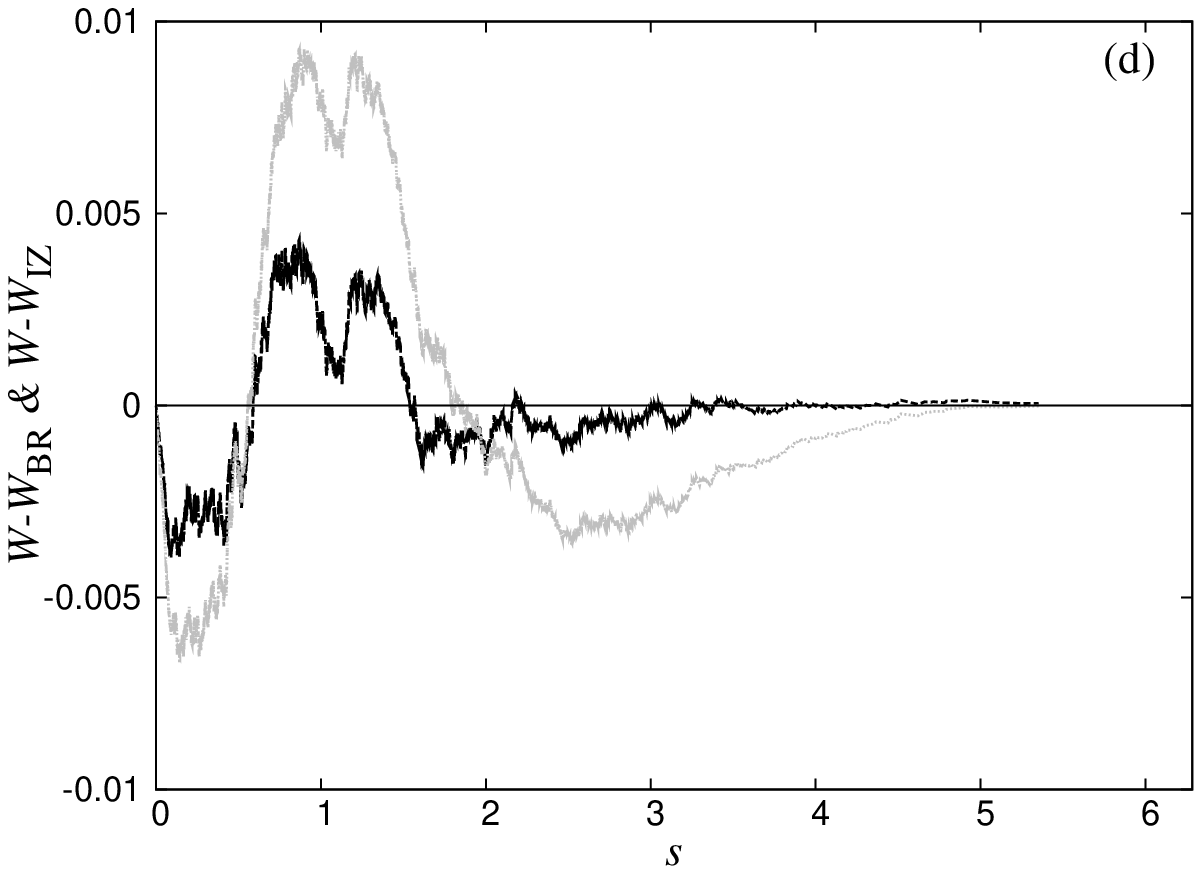}
\caption{Intermediate statistics (panel (a)) for distribution $P(s)$ (histogram - black solid line) of the model (\ref{psi_repres})-(\ref{Unm}) fitted with distribution $P_{\textrm{BR}}(s)$ (black dashed line) and $P_{\textrm{IZ}}(s)$ (black dotted line) for $N=4000$,  $K=7$ and $k=30$ (see text for discussion). The gray lines indicate the two extreme distributions, i.e. the Poisson and Wigner. In panels (b),(c),(d) we show the comparison of the the Brody (black line) and Izrailev (gray line) PDFs with the numerical using the $U$-function and $W$-distribution (see text for discussion).}
\label{pdf}
\end{figure}
Here, we use $N=4000$ (which is considerably much larger size compared to the one used in the past studies) and $K=7$ with $k=30$. In Fig.~\ref{pdf}(a) we show the numerical data (histogram) and the two PDFs. Their repulsion parameters have been calculated with best fit procedure independently. The corresponding values are found to be $\beta_{\textrm{BR}}\approx 0.424$ and $\beta_{\textrm{IZ}}\approx 0.419$ for the two PDFs respectively. In the inset figure of Fig.~\ref{pdf}(a), we may see how the $P_{\textrm{BR}}(s)$ manages to capture and describe better the peak of the distribution where the most significant part of quasienergies $\omega$ is concentrated. The dot-dashed gray line indicates the Wigner surmise while the dot-dot-dashed one the Poisson distribution. Similar findings have also been found even for smaller sizes of the matrix $U_{nm}$, where the statistics are improved by sampling more matrices with slightly different values of $k$ as e.g. in \cite{IZ_PR:1990}.

The above statement, regarding the $P_{\textrm{BR}}(s)$ better agreement with the numerical data, becomes more clear when checking the so-called $U$-functions $U(s)=(2 /\pi) \arccos \sqrt{1-W(s)}$ of the two above distributions \cite{ProRob:1994b}. The $W(s)=\int_0^sP(x)dx$ is the cumulative (or integrated) level spacing distribution function (CDF). The $U$-function has the advantage that its expected statistical error $\delta U$ is independent of $s$, being constant for each $s$ and equal to $\delta U=1/(\pi \sqrt N_s)$, where $N_s$ is the total number of objects in the $W(s)$ distribution. The numerical pre-factor $2 / \pi$ is determined in such a way that $U(s)$ $\in [0,1]$ when $W(s)$ $\in [0,1]$. We may note here that the $\beta$ values for the CDFs may be in principle slightly different compared to those found by the PDFs. In Fig.~\ref{pdf}(b), we show the $U_{\textrm BR}-U$ and $U_{\textrm IZ}-U$ vs. $U_{\textrm BR}$ where one may see that the Brody one is in general {\it closer} to zero (black line) than the Izrailev one (gray color). This fact indicates that the Brody one fits better the numerical data. This is also evident in Fig.~\ref{pdf}(c), where the $W_{\textrm BR}-W$ and $W_{\textrm IZ}-W$ vs. $W_{\textrm BR}$  are presented while in Fig.~\ref{pdf}(d) the $W_{\text BR}-W$ and $W_{\textrm IZ}-W$ vs. $s$. The horizontal zero line in these panels indicates the complete agreement between the numerical data and theoretical predictions. The repulsion parameters for the CDFs used in panels (b),(c),(d) are $\beta_{\textrm{BR}}\approx 0.396$ and $\beta_{\textrm{IZ}}\approx 0.366$ respectively.
\section{Scaling laws and localized chaotic regimes}\label{sec:scaling}

A number of different ways to measure and estimate the \textit{localization length} of the eigenfunctions have been proposed in the literature. Here, we adopt the well-accepted measure described and justified in e.g. \cite{IZ_PR:1990}: For each $N$-dimensional eigenvector of the matrix $U_{nm}$ the \textit{information entropy} is $\mathscr{H}_N(u_1,...,u_N) = -\sum_{n=1}^{N}u_n^2 \ln u_n^2$, where $u_n = {\textrm Re}\ \varphi_n$ and $\sum_n u_n^2 = 1$. The distribution of $u_n^2$ for the GOE in the large $N$-limit tends to the Gaussian distribution and we get $\mathscr{H}_{N}^{GOE}=\psi \left (0.5 N+1 \right )-\psi \left (1.5 \right )\simeq \ln \left (0.5N a \right )+O(1/N)$, where $a=4/\exp(2-\gamma)\approx 0.96$, while $\psi$ is the digamma function and $\gamma$ the Euler constant ($\simeq 0.57$). Then the {\it entropy localization length} $l_H$ is defined as $l_H=N \exp \left (\mathscr{H}_{N}-\mathscr{H}_{N}^{GOE} \right )$. The fluctuations can be minimized when using the {\it mean localization length} $<l_H> \equiv d$, which is computed by averaging over all eigenvectors of the same matrix (or over an ensemble of similar matrices) $ d=N \exp \left (<\mathscr{H}_{N}>-\mathscr{H}_{N}^{GOE} \right )$.

The parameter that determines the transition from weak to strong quantum chaos is not the strength parameter $k$ but the \textit{ratio of the localization length} $l_{\infty}$ to the size $N$ of the system, $\Lambda=l_{\infty}/N$, where $l_{\infty}=D_{cl}/2 \hbar^2$ and $D_{cl}$ is the classical diffusion constant
\begin{equation}\label{Dcl}
  D_{cl}=
    \begin{cases}
    \frac{1}{2}K^2\left [1- 2J_2(K) \left (1-J_2(K) \right ) \right ],  & \text{if} \quad K \geq 4.5 \\
    0.30(K-K_{cr})^3, & \text{if} \quad K_{cr} < K \leq 4.5
  \end{cases},
\end{equation}
where $K_{cr}\simeq 0.9716$ and $J_2(K)$ the Bessel function. The {\it localization parameter} is then defined as $\beta_{\textrm{loc}}=d/N$. The scaling law we used is $\beta_{\textrm{loc}}(x)=\gamma x/(1+\gamma x)$, where $x\equiv \Lambda$ and $\gamma \approx 4.2$ which is slightly different (but in agreement) to the one proposed in \cite{IZ_PR:1995}, where $x=k^2/N$ and $K=5$.
\begin{figure}
\centering
\includegraphics[width=6.cm]{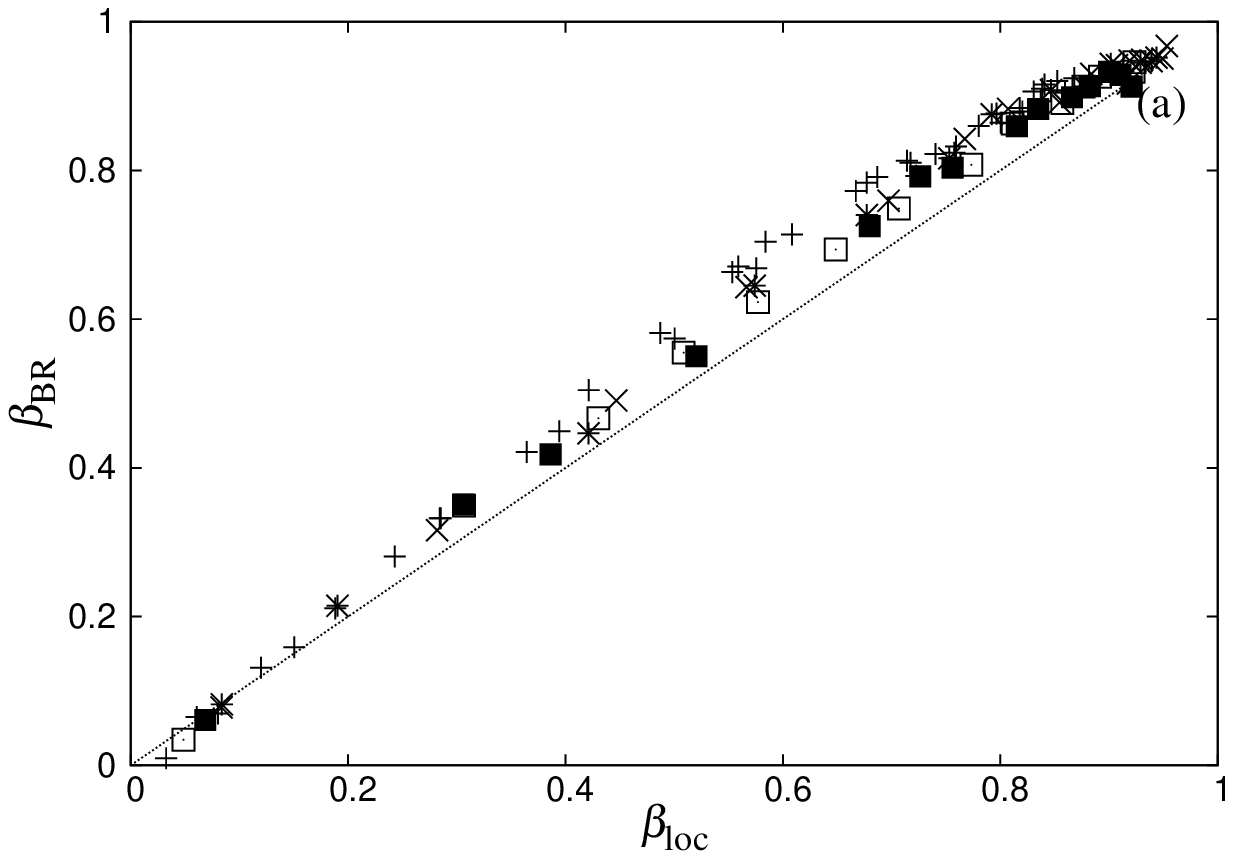}
\includegraphics[width=6.cm]{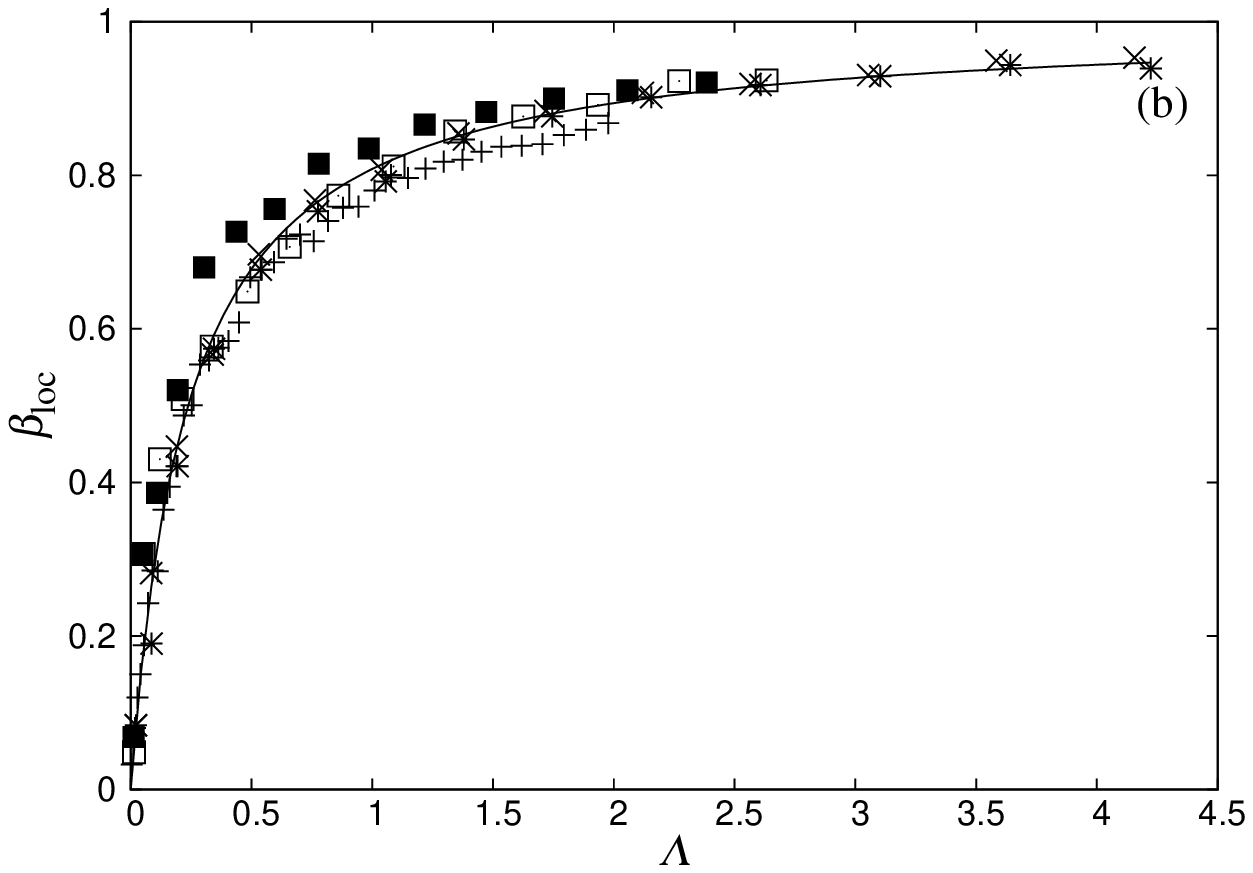}
\caption{(a) The fit parameter $\beta_{\textrm{BR}}$ as a function of $\beta_{\textrm{loc}}$ for $161\times398$ elements, $K=7(+),14(\times),20(\ast),30(\square),35(\blacksquare)$ for a wide range of $k$ values. (b) The parameter $\beta_{\textrm{loc}}$ vs. $\Lambda$ where the scaling law (see text) is shown with the black line.}
\label{betasvsL}
\end{figure}
In Fig.~\ref{betasvsL}(a), we compare $\beta_{\textrm{BR}}$ repulsion parameter of the $P_{\textrm{BR}}(s)$ with the {\it localization parameter} $\beta_{\textrm{loc}}$ through the localization length $d=<l_H>$. For the numerical calculations and results regarding the spacing distributions $P(s)$ for the quasienergies, we have considered a wide range of the quantum perturbation parameter $k$ keeping the classical parameter fixed (classically always fully chaotic). In order to ameliorate the statistics, we considered a sample of 161 matrices $U_{nm}$ of size $N=398$ ($\approx 64,000$ elements), in a similar manner as e.g. in \cite{IZ_PR:1990}) with slightly different values of $k$ ($\Delta k = \pm 0.00125 \ll k$).

\section{Summary}\label{sec:sum}
We studied aspects of dynamical localization in the kicked rotator, following \cite{IZ_PLA:1988,IZ_PR:1990,IZ_PR:1995}, and largely confirm these results. We here considered the case with $K\geq 7$, where the dynamics is already fully chaotic (ergodic). The fractional power law level repulsion is clearly manifested, and globally the level spacing distribution is very well described by the Brody or by the Izrailev distribution, with a clear and systematic (although not very large) trend towards Brody rather than Izrailev. We show that the scaling law ($\beta_{\textrm{loc}}$ vs. $\Lambda$) exists, but only as a first order approximation, as we see some scattering of data around the scaling curve. It seems that with increasing dimension of the matrices the scaling curve asymptotes to the limiting curve with only statistical
scattering of the data points left. Further research confirms that a similar picture describing  the dynamical localization applies also in time-independent systems, like e.g. billiards \cite{BatRob:2010}.

\section*{Acknowledgments}
The financial support of the Slovenian Research Agency (ARRS) is gratefully acknowledged.


\end{document}